\documentclass[12pt]{book}

\usepackage[dvips]{graphicx,color}
\usepackage{makeidx,tsukuba}

\makeauthorindex
\makeindex

\begin{document}

\BookTitle{\itshape The 28th International Cosmic Ray Conference}
\CopyRight{\copyright 2003 by Universal Academy Press, Inc.}
\pagenumbering{arabic}

\chapter{Sensitivity of the ARGO-YBJ strip size spectrum to different models of the primary cosmic ray composition in the energy range $10 \div 500~TeV$}

\author{
L. Saggese,$^1$ G. Di Sciascio,$^1$ M. Iacovacci,$^1$ S. Mastroianni,$^1$ S. Vernetto $^2$ for the ARGO-YBJ Collaboration [9]\\
{\it (1) Dip. di Fisica Universit\'a di Napoli and INFN sez. di Napoli, Italy \\
(2) IFSI-CNR and INFN, Torino, Italy}
}

\section*{Abstract}

The ARGO-YBJ experiment is currently under construction  at the Yangbajing Cosmic Ray Laboratory (4300 m a.s.l.). The detector will cover $74 \times 78~m^2$ with a single layer of Resistive Plate Counters (RPCs), surrounded by a partially instrumented guard ring. Signals from each RPC are picked-up with $80$ read out strips $6~cm$ wide and $62~cm$ long. These strips allow one to count the particle number of small size air showers. In this paper we discuss the digital response of the detector for showers with core located in a small fiducial area inside the carpet. The results enable us to assess the sensitivity of the strip size spectrum measurement to discriminate between different models of the Primary Cosmic Ray composition in the energy range $10 \div 500~TeV$.

\section{Introduction}

The energy spectrum of cosmic rays is well described by a power law over several decades of energy, before and after the so called knee region, $10^{15}~eV \div 10^{16}~eV$, where the slope changes. Despite of many conjectures and attempts, the origin of this steepening, observed in air shower data, is still obscure. Comparing existing data makes evident a substantial disagreement between the primary cosmic ray composition models provided by different experiments. As an example, the proton spectrum measured by $TIBET~AS \gamma$ [1] changes its slope at energy around $100~TeV$, $KASCADE$ [2] data suggest a steepening at about $2~PeV$ while data collected by the balloon-born experiment $RUNJOB$ [3] dont exhibit any spectral break up to $500~TeV$. In this respect, ARGO-YBJ offers an unique possibility of measuring small size air showers down to a few TeV primary energy by exploiting the digital read-out of the detector. 
In this paper we discuss the sensitivity of the ARGO-YBJ detector to discriminate between different models of the Primary Cosmic Ray composition in the energy range $10 \div 500~TeV$. 

\section{The ARGO-YBJ detector and the digital read-out}

In the ARGO-YBJ experiment (Fig.\,1 and [9]) the signals from each RPC are picked-up with $80$ read out strips $6.7~cm$ wide and $62~cm$ long for a total of $ 124800$ in the central carpet, with an average density of $\sim 22~strips/m^2$. The FAST-OR of $8$ strips defines a logic unit called PAD ($10$ PADs for each RPC) this signal being used for timing and trigger purpose. The detector is clustered in units of $12$ chambers (CLUSTER) with modular read-out (see Fig.\,1). A simulation has been carried out by means of the CORSIKA/QGSjet code [7] in order to study the dependence on the energy of the number of fired strips for quasi-vertical ($<15^o$) showers with core in a fiducial area ($A_f$) of $\sim 260~m^2$ at the center of the carpet ($2\times 3$ CLUSTER). An average strip efficiency of $95\%$ and an average strip multiplicity $m=1.2$ have been taken into account. The average strip size ($N_s$) and pad size are compared in Fig.\,2 to the total size and to the size sampled by the central carpet (truncated size) for proton induced showers. The Fig.\,2 shows clearly that the digital response of the detector can be used to study the primary spectrum up to energies of a few hundreds $TeV$.  
 \begin{figure}[t]
\vfill \begin{minipage}[t]{.40\linewidth}
 \begin{center}
 \vspace{4.pc}
   \includegraphics[height=12.5pc]{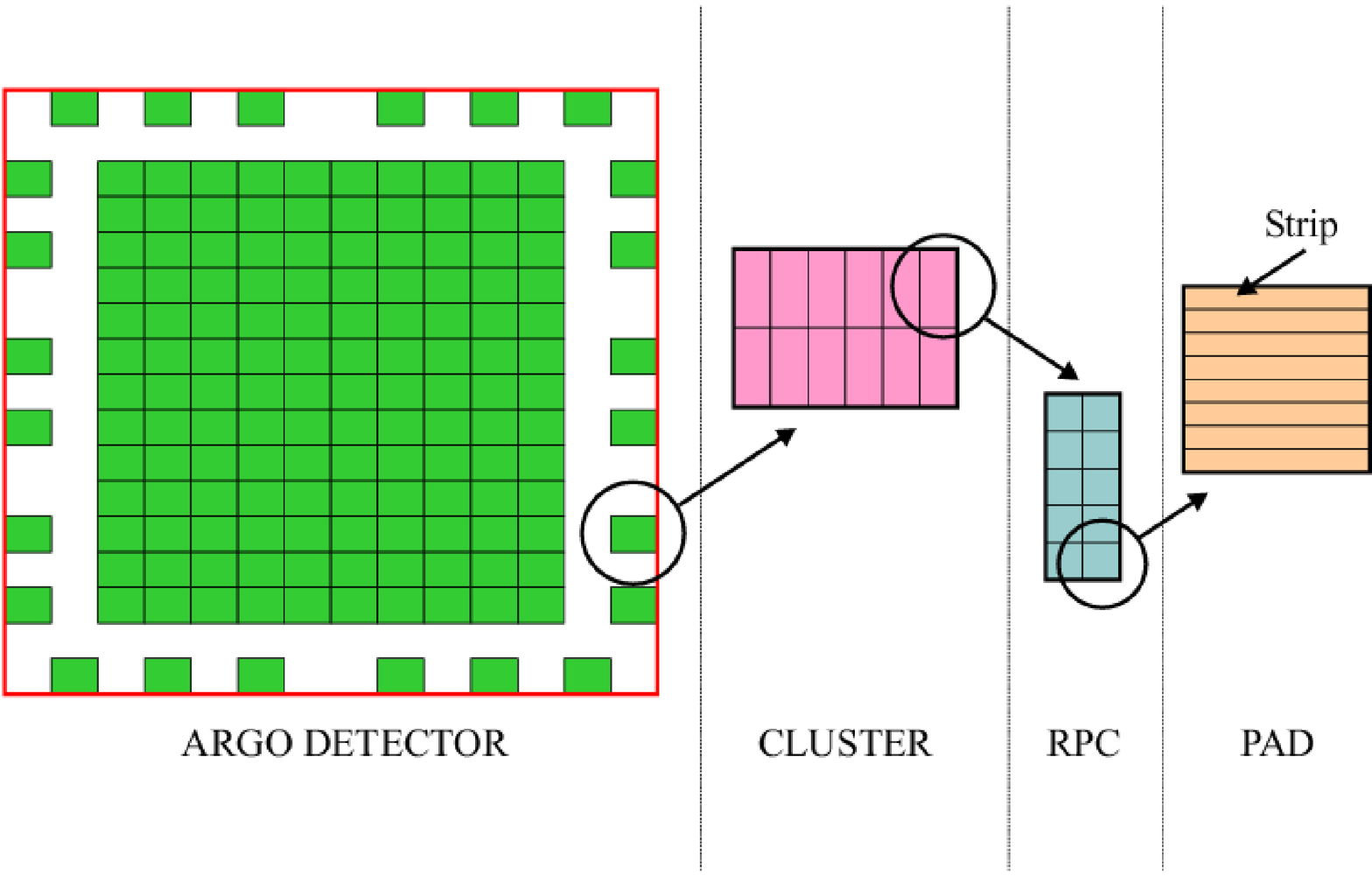}
 \end{center}
 \vspace{1.pc}
   \caption{Schematic view of the ARGO-YBJ detector .}
\end{minipage}\hfill
\hspace{-0.5cm}
\begin{minipage}[t]{.45\linewidth}
 \begin{center}
   \includegraphics[height=17.5pc]{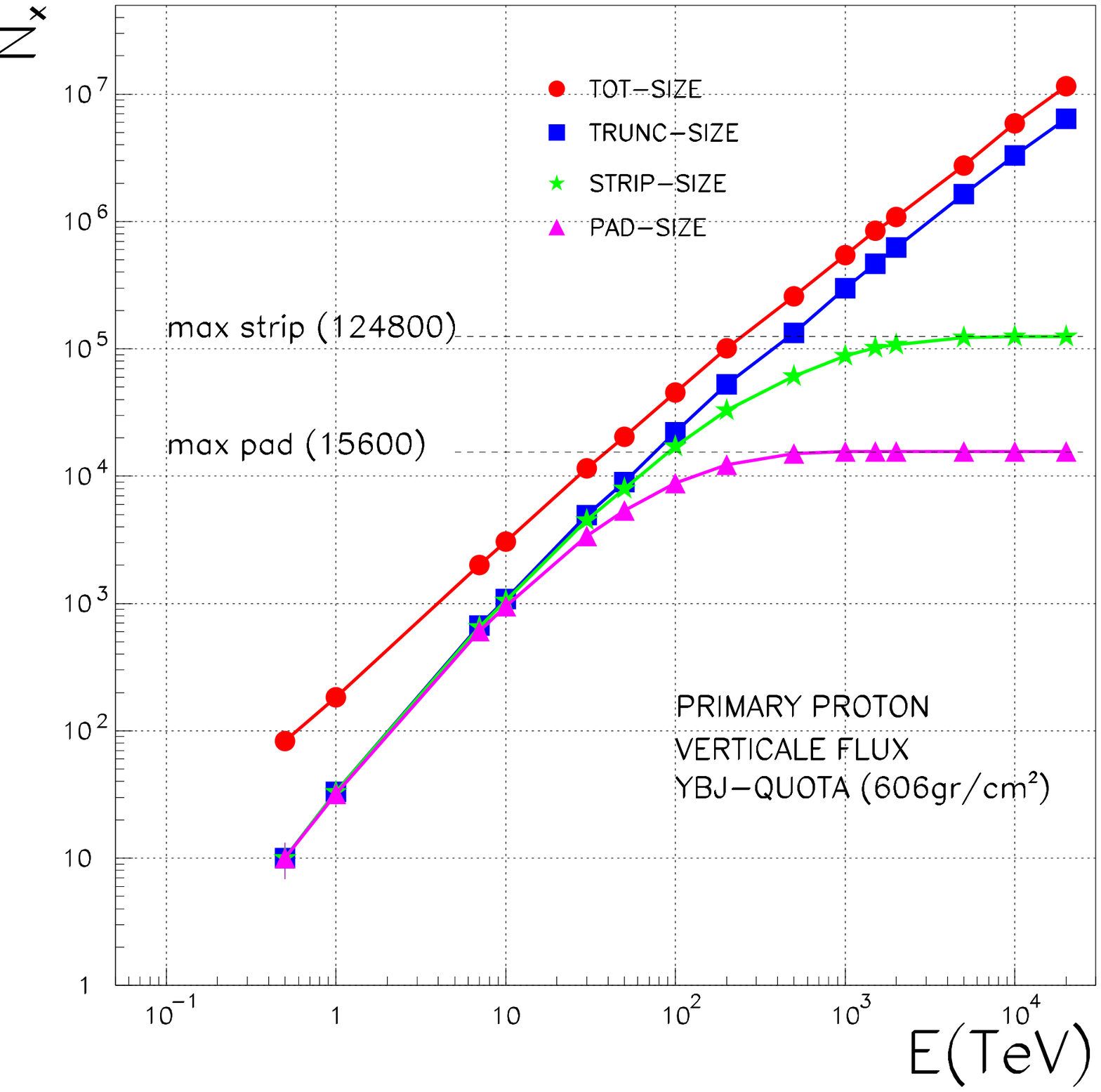}
 \end{center}
 \vspace{-1.5pc}
   \caption{Average strip size and pad size compared to the  total size and truncated size .}
\end{minipage}\hfill
\end{figure}

\section{The strip size spectrum}

To check the sensitivity of a digital measurement to the primary spectrum, the strip size spectrum has been obtained for four models of the primary cosmic ray composition by calculating the differential flux  
\begin{equation}\label{all-siz1}
\frac{dN}{dy_s} \equiv J_{all}(y_s)=\sum_{A} \varepsilon_A (y_s) \cdot \int_{E_{min}}^{\infty} J_{A} (E) \cdot P_{A}(y_s,E) \cdot dE
\end{equation}

\noindent where $y_s=Log(N_s)$, $J_{A} (E)$ is the differential intensity for each nuclear species, $P_{A}(y_s,E)$ is the probability that a nucleus of mass number $A$ and energy $E$ produces a strip size $y_s$. The following models for the primary composition have been considered: $JACEE$ [4]; $RUNJOB$ [3]; $TIBET~AS \gamma$ [1] and the model proposed by $J.~ R.~ Horandel$ [6] to fit world data from EAS experiments. To get $P_{A}(y_s,E)$ we have envisaged the following experimental conditions: 1) a trigger requiring at last $16$ PADs out of $120$ in any one of the $4\times 5$ central CLUSTER ($\sim 1000~m^2$) including the fiducial area $A_f$; 2) shower core and shower direction reconstructed following the algorithms developed in [5] and [8] respectively. The probalilities for shower with core recostructed in $A_{f}$ and zenith angle $\theta < 15^o$ may be described by a gaussian form $P_{A}(y_s,E)=\frac {1}{\sqrt{2 \pi} \cdot \sigma_{y_s}(A,E)} \cdot exp [- \frac{1}{2} \cdot [\frac{y_{s}- \overline {y}_{s}(A,E)}{\sigma_{y_{s}}(A,E)}]^{2}]$. Showers which dont satisfy the trigger condition or are reconstructed with core outside the fiducial area are not considered. The efficiency $\varepsilon_A (y_s$) , see Fig.\,3, represents the fraction of triggered and reconstructed events with core in $A_f$. In Fig.\,4 is plotted the function $F_{all}(y_s)=J_{all}(y_s)\cdot 10^{1.5 \cdot y_s} \cdot \Gamma \cdot \Delta y_s$
\noindent with $\Delta y_s=0.1$ and $\Gamma=10^8~m^2ssr$. This function provides in a simple way the slope of the strip size spectrum ($J_{all}(N_s) \propto N_s^{-\alpha_{s}}$) in the interval $y_{s_{1}} \div y_{s_{2}}$, $\alpha_{s} = 2.5 - \frac{Log[F_{all}(y_{s_2})/F_{all}(y_{s_1})]}{y_{s_{2}}-y_{s_{1}}}$
\noindent and the counting rate number of collected events, $C_s (y_s)$, integrated in $\Delta y_s=0.1$ for an exposure $\Gamma=10^8~m^2ssr$, corresponding to about one month of data taking for showers with core selected in $A_f$ and zenith angle $\theta < 15^o$, $C_s (y_s)= \frac {F_{all}(y_{s})}{10^{1.5 \cdot y_s}}$.  
 \begin{figure}[t]
\vfill \begin{minipage}[t]{.48\linewidth}
 \begin{center}
   \includegraphics[height=17.pc]{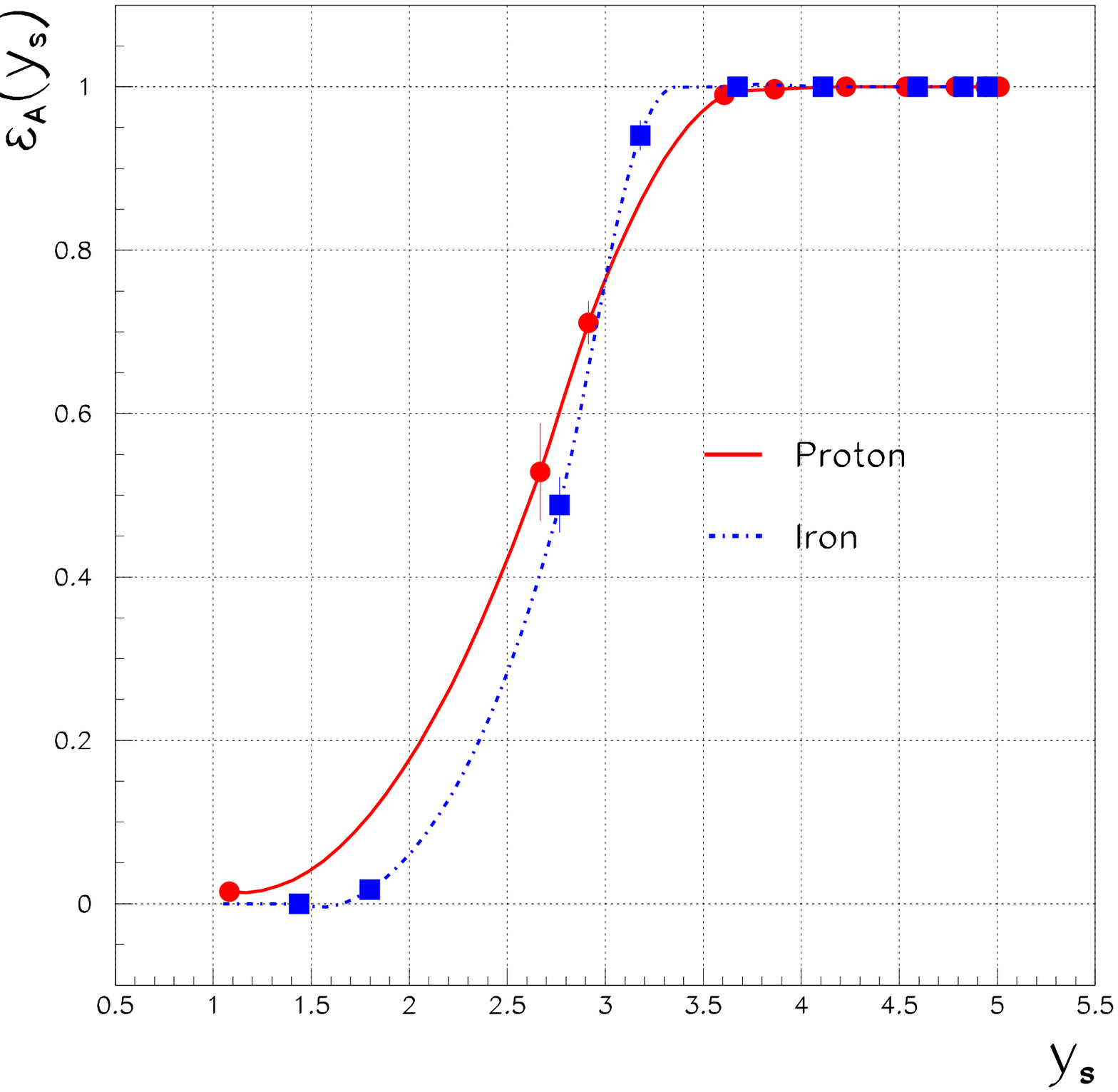}
 \end{center}
 \vspace{-1.5pc}
   \caption{Fraction of triggered and recostructed events with core in the fiducial area for proton and iron induced air shower.}
\end{minipage}\hfill
\hspace{-1.cm}
\begin{minipage}[t]{.48\linewidth}
 \begin{center}
   \includegraphics[height=17.pc]{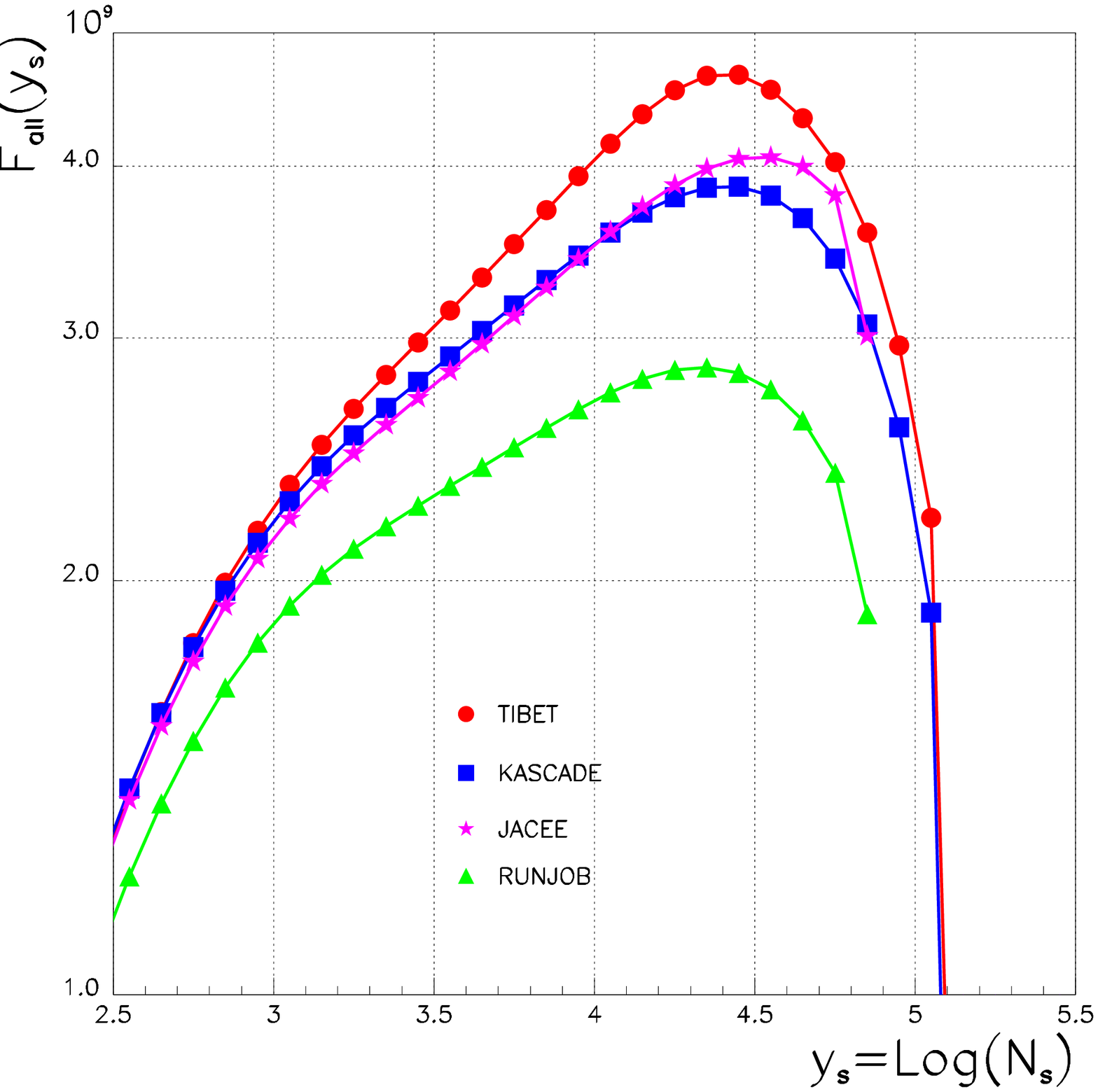}
 \end{center}
 \vspace{-1.5pc}
   \caption{Size strip spectrum for four models of primary composition (see text).}
\end{minipage}\hfill
\end{figure}

\section{Discussion and conclusions}

From the results of Fig.\,4 we can draw the following conclusions: 

1. the strip size spectra pile-up above $y_s \simeq 4.2$ ($<E_p> \simeq 100~TeV$, $<E_{Fe}> \simeq 320~TeV$), due to the saturation of the digital read-out, and fall down below $y_s \simeq 3$ ($<E_p> \simeq 10~TeV$, $<E_{Fe}> \simeq 40~TeV$) owing to a substantial decrease of the trigger efficiency ($\varepsilon (y_s=3) \simeq 75\%$, see Fig.\,4); 

2. in the range $y_s=3 \div 4$ the $JACEE$ data and the {\em Horandel model} predict the same strip size spectrum with $\alpha_s \simeq 2.35$. This is mainly due to the fact that in these models the spectrum of light component is quite similar;

3. the biggest difference is found between the strip size spectra obtained using the $TIBET~AS \gamma$ and $RUNJOB$ models. The spectral index is $\simeq 2.27$ and $\simeq 2.37$, respectively, and the counting rate expected according to the $TIBET~AS \gamma$ spectrum is higher than the one predicted by the $RUNJOB$ data, running from $\sim 30\%$ at $y_s \simeq 3$ up to $\sim 55\%$ at $y_s \simeq 4$;    

4. the number of events in each size bin is enough to make negligible the statistical uncertainty. On the contrary, any systematic error $\delta N_s/N_s$ in reconstructing the strip size spectrum determines a shift $\Delta F_{all}/F_{all}=(\delta N_s/N_s +1)^{\alpha_s -1} -1$. Thus, a control of the detector performance at a level better than $10\%$ is required in order to reduce any systematic effect below $15\%$;   

5. the spectra expected from using a different hadronic model have not been calculated. Due to low energy range involved, we expect differences lower than $10\%$. 

The possibility of discriminating between different models of the primary cosmic ray composition by using the digital read-out of ARGO-YBJ has been investigated. It is important to notice that this technique should allow to scan the energy range from $10$ to a few hundred $TeV$ where direct and indirect measurements partially overlap.    

\vspace{1.8pc}

\re
1. Amenomori M. et al.\ 2001, Phys.Rev. D, 62, 112001-1.
\re
2. Antoni T. et al.\ 2002, Astr.Phys. 16, 373.
\re
3. Apanasenko A. V. et al.\ 2001, Astr.Phys. 16, 13.
\re
4. Asakimori R. et al.\ 1998, ApJ 502, 278.
\re
5. Di Sciascio G. et al. (ARGO-YBJ Coll.) in this proceedings.
\re
6. Horandel J. R. et al. \ 2001, Proc. 27th ICRC Hamburg, 1, 71.
\re
7. Knapp J. et al.\ 1998, Forschungszentrum Karls\-ruhe  FZKA 6019.
\re
8. Martello D. et al. (ARGO-YBJ Coll.) in this proceedings.
\re
9. Surdo A. et al. (ARGO-YBJ Coll.) in this proceedings.

\endofpaper
\end{document}